\documentclass[prl,reprint,showpacs,showkeys,amsmath,amssymb,aps]{revtex4-1}

\usepackage{epsfig}
\usepackage{graphicx}%
\usepackage{dcolumn}
\usepackage{bm}

\def\2pap{2\pi\alpha^\prime}

\def\beq{\begin{eqnarray}}
\def\eeq{\end{eqnarray}}

\begin{document}

\title[Short Title]{Understanding the Complex Position in a $\mathcal{PT}$-symmetric Oscillator}

\author{Jin-Ho Cho}\email{jinhocho@hanyang.ac.kr} 
 \affiliation{Department of Physics \& Research Institute for Natural Sciences, Hanyang University, Haengdang-dong, Seongdong-gu, Seoul 133-791, Korea} 

\date{\today}

\begin{abstract}
We study how to understand the complex coordinates involved in the non-Hermitian but $\mathcal{PT}$-symmetric systems. We explore a $\mathcal{PT}$-symmetric oscillator model to show that the entire information on the complex position is attainable. Its real part is from the observation while its imaginary part is from the non-Hermiticity parameter. We also propose a new complex extension of $\mathcal{P}$-transformation and $\mathcal{T}$-transformation (the `parity' and `time reflection' respectively). Particularly, the $\mathcal{P}$-transformation realizes the left-right reflection in the complex plane.
\end{abstract}

\maketitle
{Solving the eigenvalue problem of a system associated with `non-Hermitian' potential, we generically meet the complex coordinates. For example, in the non-Hermitian potential like $V(x)\sim x^2(ix)^\nu$ proposed in Ref. \cite{bender}, the turning points, which are relevant in finding the connection formula in WKB approximation, are determined by the relation $E=V(x)$. The points are definitely in the complex plane. }

{It is confusing that we have to work in complex coordinates finding the connection formula in WKB method. We already know that the position operator $\hat{x}$ is Hermitian with respect to the $L^2$-inner product (the conventional one we use in quantum mechanics), thus has real eigenvalues.}

{The aim of this letter is to clarify this confusion by inspecting a `non-Hermitian' but $\mathcal{PT}$-symmetric harmonic oscillator. (We will be more specific about these jargons later.)}

{As a warmup for the non-Hermitian system, let us consider a massive charged particle, living in one-dimension under an external potential $V(x)$, and being coupled with an external constant `gauge' $A$; 
\begin{equation}\label{hamil}
\hat{H}\psi(x)=\left\{ \frac{1}{2}\left(\hat{p}-\hat{A}\right)^2+\hat{V}(\hat{x})\right\}\psi(x).
\end{equation}
Since no electric field or magnetic field is involved, one can neglect the trivial gauge coupling by redefining the wave function as $\psi(x)=e^{iAx}\phi(x)$}. 

{What if the constant `gauge' $A$ is pure imaginary? Indeed, one occasionally encounters such a case in the solid state physics system exhibiting the delocalization phenomena \cite{nelson} or in the optics system like the channel wave guide involving the surface wave in the cladding layers \cite{kitoh}. The above method of redefining the wave function is no help if $A$ is pure imaginary. Though the newly defined wave function $\phi(x)$ is normalizable, the function $\psi(x)$ diverges in the asymptotic region. The difficulty with the case of the imaginary $A$ is originating from the non-Hermiticity of the Hamiltonian resulting in non-unitary time evolution. } 

{A class of non-Hermitian Hamiltonians can possess a real spectrum. Behind those systems lies $\mathcal{PT}$-symmetry \cite{bender}, or pseudo-Hermiticity \cite{mostafazadeh}. In a broader sense, $\mathcal{P}$- and $\mathcal{T}$-transformation are a linear and an anti-linear involution on the phase space preserving the commutation relation $\left[\hat{x},\,\hat{p} \right]=i\hat{I}$ \cite{Mostafazadeh:2002pd}. Under the combined action of $\mathcal{P}$ ($\hat{x} \rightarrow -\hat{x}$, $\hat{p} \rightarrow -\hat{p}$, $i\hat{I} \rightarrow i\hat{I}$) and $\mathcal{T}$ ($\hat{x} \rightarrow \hat{x}$, $\hat{p} \rightarrow -\hat{p}$, $i\hat{I} \rightarrow -i\hat{I}$), the above Hamiltonian (with the imaginary $A$) will be invariant if $\hat{V}^{\mathcal{PT}}(\hat{x})=\hat{V}(\hat{x})$.} 

{A concrete model in the class was first provided and pioneered in Refs. \cite{bender,Bender:2002vv,Bender:2007nj}, where $\hat{A}=0$ and $\hat{V}(\hat{x})=\hat{x}^2(i\hat{x})^\nu$, thus is $\mathcal{PT}$-symmetric. It was shown by numerical method or WKB approximation that the system with $\nu\geq 0$ has real eigenvalues when $\mathcal{PT}$-symmetry is exact, that is when $\mathcal{PT}$ shares its eigenstates with the Hamiltonian. Experimental observation on the symmetry has been put forward in optics \cite{ruter}, especially in the wave guide physics. (See Ref. \cite{hanyang} and references therein.)}

{In this letter, we consider a model with 
\begin{equation}
\hat{A}=iz^*\hat{I},\qquad \hat{V}(\hat{x})=\frac{1}{2}\left( \hat{x}-z^*\hat{I}\right)^2,
\end{equation}
where $\hat{I}$ is the identity operator and $z^*=\rho \,e^{i\lambda}$ is a complex number. Unless $z$ is pure imaginary ($\lambda=\pm \pi/2$), $\mathcal{PT}$ symmetry is not clear. At the final stage, we will see the system has a real spectrum for arbitrary value of $z$ and how $\mathcal{PT}$ symmetry reconcile with the general case.}

{The model has the virtue of exact solvability. Moreover, it has not only the coordinate space representation but also the spectral representation, which allows a clear connection between the $\mathcal{PT}$ symmetry (studied in Refs. \cite{bender,Bender:2002vv}) and the pseudo-Hermiticity (discussed in Refs. \cite{mostafazadeh,Mostafazadeh:2008pw}). }

{Regarding the spectral representation, it is more instructive to write the Hamiltonian in terms of the laddering operators; 
\begin{equation}\label{repre}
\hat{a}=\left(\hat{x}+i\hat{p} \right)/\sqrt{2},\quad \hat{a}^\dagger=\left(\hat{x}-i\hat{p} \right)/\sqrt{2}
\end{equation}
(Note that we have set $m=\hbar=\omega=e=1$ for simplicity.)
The non-Hermiticity is clear in this representation, 
\begin{equation}\label{nonhermitianH}
\hat{H}=\left(\hat{a}^\dagger-z^*\sqrt{2}\hat{I} \right)\hat{a}+\frac{1}{2}.
\end{equation}
We hope it to be pseudo-Hermitian instead, because it will have a real spectrum then. (See Ref. \cite{mostafazadeh} for the rigorous mathematics behind the pseudo-Hermitian system.) The form suggests that the laddering operators be redefined as 
\begin{equation}\label{ladderb}
\hat{b}^\sharp:=e^{-i\theta}(\hat{a}^\dagger-z^*\sqrt{2}\hat{I}),\qquad \hat{b}:=\hat{a}e^{i\theta}.
\end{equation}
Indeed, they compose laddering operators satisfying $[\hat{b},\,\hat{b}^\sharp ]=\hat{I}$. If $\hat{b}^\sharp$ is the pseudo-Hermitian conjugate of $\hat{b}$, more specifically if $\hat{b}^\sharp=\eta^{-1}\hat{b}^\dagger\eta$ for some Hermitian operator $\eta$, the Hamiltonian will be pseudo-Hermitian, that is, $\hat{H}=\eta^{-1}\hat{H}^\dagger\eta=\hat{H}^\sharp$. Inspired by the coherent state computation, one can determine it as $\eta=e^{z^*\sqrt{2}\hat{a}+z\sqrt{2}\hat{a}^\dagger}$. It is not only Hermitian but also positive in the sense that $\langle \psi\vert \eta\vert \psi\rangle>0$ for an arbitrary state $\vert \psi\rangle$. In the case, the operator $\eta$ is termed the `metric operator' and the pseudo-Hermitian Hamiltonian is more specified to be quasi-Hermitian.}

{Two different Fock spaces are involved in the system. On the common vacuum $\vert 0\rangle_a=\vert 0\rangle_b:=\vert 0\rangle$ (annihilated by $\hat{b}=\hat{a}e^{i\theta}$), one can successively apply either $\hat{a}^\dagger$ or $\hat{b}^\sharp$ to construct the Fock space $\{\vert n\rangle_a\}$ or $\{\vert n\rangle_b\}$. Their relation is
\begin{equation}\label{abrel}
\vert n\rangle_b=e^{-in\theta}\sum^n_{l=0}\sqrt{{}_nC_l}\frac{(-z^*\sqrt{2})^l}{\sqrt{l!}}\vert n-l\rangle_a.
\end{equation}
Only the states $\{\vert n\rangle_b\}$ form the eigenstates because the hamiltonian is proportional to the number operator $\hat{N}_b=\hat{b}^\sharp \hat{b}$ associated with the set $\{\hat{b},\,\hat{b}^\sharp\}$.}

{The eigenstates $\vert n\rangle_b$ of the quasi-Hermitian Hamiltonian $\hat{H}$ are not orthonormal with respect to $L^2$-inner product. Indeed, an explicit computation shows that
\begin{equation}
{}_b\langle m\vert n\rangle_b=\mathcal{N}_{mn}\!\!\!\sum^{\text{min}\left\{m, n\right\}}_{k=0}\!\!\!{}_mC_{k}\, {}_nP_{k}\frac{(z)^{m}(z^*)^{n}}{\left\vert z\right\vert^{2k}}
\end{equation}
where the numerical factor is determined as $\mathcal{N}_{mn}=e^{i(m-n)\theta}(-\sqrt{2})^{m+n}/\sqrt{m!n!}$.}

{The `non-Hermitian' Hamiltonian \eqref{nonhermitianH} is Hermitian with respect to a new inner-product $\langle\cdot\vert\cdot\rangle_\eta:=\langle\cdot\vert\eta\vert\cdot\rangle$ (hereafter called as $\eta$-inner product). Indeed $\langle\phi\vert \hat{H}\psi\rangle_\eta\!=\!\langle\phi\vert\eta \hat{H}\psi\rangle\!=\!\langle\hat{H}^\sharp\phi\vert \psi\rangle_\eta\!=\!\langle\hat{H}\phi\vert \psi\rangle_\eta$ for arbitrary states $\vert\phi\rangle$ and $\vert\psi\rangle$. Specifically for the choice of two eigenstates $\vert m\rangle_b$ and $\vert n\rangle_b$, it implies that the eigenvalues are real, that is, $E_m=E^*_m$ and the eigenstates can be orthonormal, ${}_b\langle m\vert \eta\vert n\rangle_b=\delta_{mn}$, with respect to the $\eta$-inner product.}

{The Hamiltonian has not only the real spectrum but also the real expectation values. According to Dirac-von Neumann axiom of quantum mechanics, the expectation value rather than the spectrum concerns the observation \cite{dirac,neumann}. An observable corresponds to an operator in the Hilbert space and it is Hermitian if and only if it has real expectation values \cite{Mostafazadeh:2008pw}. The Hamiltonian \eqref{nonhermitianH}, though non-Hermitian with respect to $L^2$-inner product, can possess real expectation values as far as $\eta$-inner product is used. The Hamiltonian \eqref{nonhermitianH} can delineate a physical system if $\eta$-inner product is adopted. } 

{The problem with using $L^2$-inner product for the system \eqref{nonhermitianH} is that the energy expectation value for a non-eigenstate depends on time, and what is worse, it develops imaginary value as time flows, thus is unphysical in the sense of Dirac-von Neumann axiom. Indeed for a specific state $\vert \psi\rangle=\vert 1\rangle_a=z^*\sqrt{2}\vert 0\rangle_b+e^{i\theta}\vert 1\rangle_b$, the energy expectation value with respect to $L^2$-inner product is given by
\begin{equation}
\frac{\langle\psi_t\vert\hat{H}\vert\psi_t\rangle}{\langle \psi_t\vert\psi_t\rangle}=1+\frac{1+4i \left\vert z\right\vert^2 \sin{t}\,}{2+8 \left\vert z\right\vert^2 \left( 1-\cos{t}\,\right)},
\end{equation}
whereas with respect to $\eta$-inner product it is a real constant;
\begin{equation}
\frac{\langle\psi_t\vert\hat{H}\vert\psi_t\rangle_\eta}{\langle\psi_t\vert\psi_t\rangle_\eta}=\frac{3+2\left\vert z\right\vert^2}{2+4\left\vert z\right\vert^2}.
\end{equation}
The results are illustrated in Fig. \ref{expecr}. (The figures are generated by using Mathematica \cite{mathematica}.)
\begin{figure}[htbp]
\centering
\includegraphics[width=6cm]{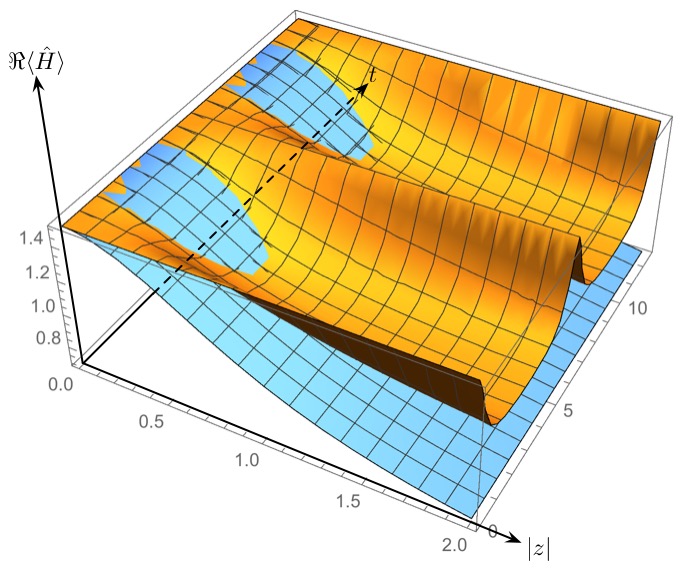}
\includegraphics[width=6cm]{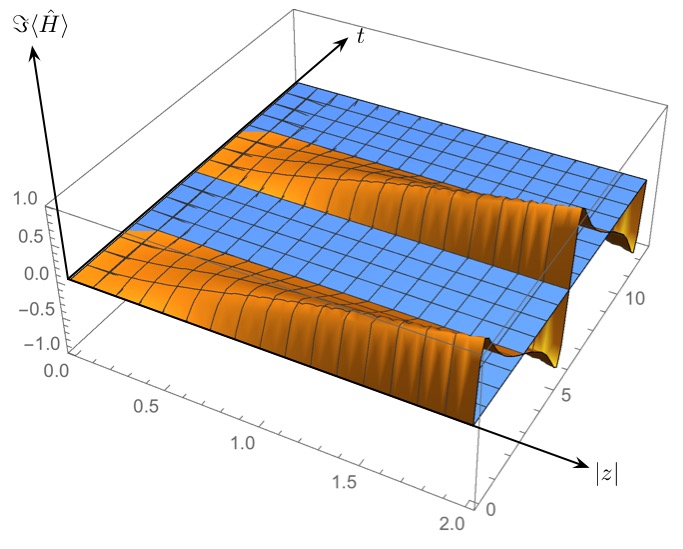}
\caption{\small The energy expectation value for a non-eigenstate depends on the inner product used in the computation. The yellow surface concerns $L^2$-inner product $\langle\cdot\vert\cdot \rangle$ while the blue is associated with $\eta$-inner product $\langle\cdot\vert\eta\cdot \rangle$. The upper figure shows the real part of the expectation value and the lower figure displays the imaginary part. When $\left\vert z\right\vert\ne 0$, the values in the $L^2$-inner product are periodic in time. On the while, no imaginary part develops in the $\eta$-inner product. The case when $\left\vert z\right\vert=0$ corresponds to the conventional harmonic oscillator.}
\label{expecr}
\end{figure}
}

{The orthonormality condition ${}_b\langle m\vert \eta\vert n\rangle_b=\delta_{mn}$ suggests that the states $\vert n\rangle_{b'}:=\eta\vert n\rangle_b$ are dual to the states $\vert m\rangle_b$. They are the eigenstates of the operator $\hat{H}^\dagger=\hat{b}^\dagger\hat{b}^{\sharp\dagger}+1/2$ satisfying $\hat{H}^\dagger\vert n\rangle_{b'}=E^*_n\vert n\rangle_{b'}$, and compose the biorthonormal set of eigenstates along with the states $\vert m\rangle_b$. Note that the operators $\hat{b}^\dagger$ and $\hat{b}':=\hat{b}^{\sharp\dagger}=\eta\hat{b}\eta^{-1}$ form the dual set of laddering operators satisfying $[ \hat{b}',\hat{b}^\dagger]=\hat{I}^\dagger$.}

{The biorthonormal set of eigenstates allows the spectral representations for various operators. Some of them are $\hat{H}=\sum^\infty_{n=0}\!E_n\vert n\rangle_{b}\,{}_{b'}\!\langle n\vert$ and $\eta=\sum^\infty_{n=0}\!\vert n\rangle_{b'}\,{}_{b'}\!\langle n\vert$.
It is easy to see that they satisfy the pseudo-Hermiticity condition $\hat{H}=\hat{H}^\sharp$.
}

{Let us find the anti-linear involution symmetry associated with the system for arbitrary value of $z$. A pseudo-Hermitian Hamiltonian admits an anti-linear involution operator as a symmetry \cite{Mostafazadeh:2002pd}. As was mentioned earlier, $\mathcal{PT}$ symmetry looks clear when $z=\pm i\rho$. A systematic way of constructing the generalized $\mathcal{P}$-, $\mathcal{T}$-, and $\mathcal{C}$-operator has been developed in Ref. \cite{Mostafazadeh:2002pd}, which we will employ here dealing with the case of arbitrary complex $z$. Associated with the metric operator $\eta$, there are canonical pseudo-metric operators and the anti-linear operators
\begin{equation}
\mathcal{P}_\sigma\!=\!\sum^\infty_{n=0}\sigma_n\vert n\rangle_{b'}{}_{b'}\langle n\vert,\quad
\mathcal{T}_{\sigma'}\!=\!\sum^\infty_{n=0}\sigma'_n\vert n\rangle_{b'}\star{}_{b'}\langle n\vert.\nonumber
\end{equation}
Here, the series elements $\sigma_n$ and $\sigma'_n$ take values in $\left\{-1,\,1 \right\}$ and the operation $\star$ is the complex conjugation on the expression appearing on its right. With proper normalization of the metric operator as $\eta \rightarrow\tilde{\eta}=e^{-2 \left\vert z\right\vert^2}\eta$, one can show that the operator $\mathcal{P}_{\sigma}$  with $\sigma_n=(-1)^n$ is an involution, that is, it is squared to the identity. The operator $\mathcal{T}_{\sigma'}$ can be an involution too. The sigma factor is identified as $\sigma'_m=(-1)^m$ when $\theta-\lambda=0,\,\pi,\,2\pi,\cdots$, and while $\sigma'_m=1$ if $\theta-\lambda=\pi/2,\,3\pi/2,\,5\pi/2,\cdots$. This means that for arbitrary complex value of $z^*=\rho e^{i\lambda}$, one can always adjust the phase $\theta$ in $\hat{b}=\hat{a}e^{i\theta}$ so that $\mathcal{P}_\sigma$ and $\mathcal{T}_{\sigma'}$ be involutions.}

{From the above involutive operators, one can construct involutive symmetry generators as $\mathcal{C}_\sigma:=\tilde{\eta}^{-1}\tilde{\eta}_\sigma$, and $\mathcal{X}_{\sigma\sigma'}:=\mathcal{P}^{-1}_\sigma\mathcal{T}_{\sigma'}=\mathcal{P}_\sigma\mathcal{T}_{\sigma'}$. They commute with the Hamiltonian because they satisfy $\hat{H}^\dagger=\tilde{\eta}\hat{H}\tilde{\eta}^{-1}=\mathcal{P}_\sigma\hat{H}\mathcal{P}^{-1}_\sigma=\mathcal{T}_\sigma\hat{H}\mathcal{T}^{-1}_\sigma$. Their spectral representations are $\mathcal{C}_\sigma=\sum^\infty_{n=0}\sigma_n\vert n\rangle_b\,{}_{b'}\langle n\vert,$ and $\mathcal{X}_{\sigma\sigma'}=\sum^\infty_{n=0}\sigma_n\sigma'_{n}\vert n\rangle_{b}\star {}_{b'}\langle n\vert.$}

{The spectral representation, 
\begin{eqnarray}
b&=&\sum^\infty_{n=0}\sqrt{n+1}\vert n\rangle_b\,{}_{b'}\langle n+1\vert, \nonumber\\
b^\sharp&=&\sum^\infty_{n=0}\sqrt{n+1}\vert n+1\rangle_b\,{}_{b'}\langle n\vert 
\end{eqnarray}
enables us to compute the following transformations explicitly. Under the pseudo-metric operator $\mathcal{P}_\sigma$, $\hat{b}\rightarrow -\hat{b}'$, $\hat{b}^\sharp\rightarrow -\hat{b}^\dagger$, and $i\hat{I}\rightarrow i\hat{I}'$. Here $\hat{I}'=\hat{I}^\dagger$ represents the identity operator on $b'$-Fock space. Under the anti-linear operator $\mathcal{T}_{\sigma'}$, $\hat{b}\rightarrow \mp\hat{b}'$, $\hat{b}^\sharp\rightarrow \mp\hat{b}^\dagger$, and $i\hat{I}\rightarrow -i\hat{I}'$. The upper/lower sign is for the case when $\theta-\lambda$ is an integer/a half-integer multiple of $\pi$, respectively. (The same correspondence will be applied upon the appearance of the multiple signs hereafter.) Lastly under the generalized charge operator $\mathcal{C}_\sigma$, $\hat{b}\rightarrow -\hat{b}$, $\hat{b}^\sharp\rightarrow -\hat{b}^\sharp$, and $i\hat{I}\rightarrow -i\hat{I}$.}

{The phase space variables $\hat{x}$ and $\hat{p}$ transform nontrivially under the above involutions. Exploiting the relations \eqref{repre} and \eqref{ladderb}, we obtain
\begin{eqnarray}
\mathcal{P}_\sigma:&&\,\hat{x}\rightarrow-\hat{x}+(z+z^*)\hat{I}', \qquad\hat{p}\rightarrow -\hat{p}-i(z-z^*)\hat{I}',  \nonumber\\
\mathcal{T}_{\sigma'}:&&\,\hat{x}\rightarrow\mp(\hat{x}\cos{2\theta}\,-\hat{p}\sin{2\theta}\,)+(z+z^*)\hat{I}',  \nonumber\\
&&\,\hat{p}\rightarrow\pm(\hat{x}\sin{2\theta}\,+\hat{p}\cos{2\theta}\,)-i(z-z^*)\hat{I}',\nonumber\\
\mathcal{C}_\sigma:&&\,\hat{x}\rightarrow-\hat{x}+2z^*\hat{I},\qquad\hat{p}\rightarrow-\hat{p}+2iz^*\hat{I}.\nonumber
\end{eqnarray} They look entirely different from the ones given in Ref. \cite{bender}, or those mentioned in the earlier part of this letter. Under the combination of $\mathcal{P}_\sigma\mathcal{T}_{\sigma'}$, the operators transform homogeneously as $\hat{q}\rightarrow\pm(\hat{x}\cos{2\theta}\,-\hat{p}\sin{2\theta}\,)$, $\hat{p}\rightarrow\mp(\hat{x}\sin{2\theta}\,+\hat{p}\cos{2\theta}\,)$. Only with $z$ pure imaginary, thus $\cos{2\theta}\,=\mp1$, the result accords with those in Ref. \cite{bender}.}

{It is possible to define the pseudo-Hermitian phase variables. By construction, the operators, $\hat{X}:=(\hat{b}+\hat{b}^\sharp)/\sqrt{2}$ and $\hat{P}:=-i(\hat{b}-\hat{b}^\sharp)/\sqrt{2}$ are pseudo-Hermitian possessing real spectra. They are physical in the sense that their expectation values are real. They transform homogeneously under the involutions:
\begin{eqnarray}
\mathcal{P}_\sigma:&&\!\!\!\hat{X}\rightarrow-\hat{X}^\dagger, \quad\hat{P}\rightarrow -\hat{P}^\dagger,\quad i\hat{I}\rightarrow i\hat{I}',  \nonumber\\
\mathcal{T}_{\sigma'}:&&\!\!\!\hat{X}\rightarrow\mp\hat{X}^\dagger, \quad\hat{P}\rightarrow\pm\hat{P}^\dagger,\quad i\hat{I}\rightarrow -i\hat{I}',\nonumber\\
\mathcal{C}_\sigma:&&\!\!\!\hat{X}\rightarrow-\hat{X},\quad\hat{P}\rightarrow-\hat{P},\quad i\hat{I}\rightarrow -i\hat{I}.
\end{eqnarray}
When $\theta-\lambda$ is a half-integer multiple of $\pi$, $\sigma'_m=1$ and the transformations (with lower sign) look analogous to the one given earlier. However, the transformations relate the operators $\hat{X}, \hat{P}$ with their Hermitian conjugates $\hat{X}^\dagger, \hat{P}^\dagger$. They amount to the complex extension of $\mathcal{P}$- and $\mathcal{T}$-transformation.  A different complex extension of $\mathcal{P}$- and $\mathcal{T}$-transformation is given in Ref. \cite{Bender:1998gh}, but we emphasize that the above transformation is perfectly allowable too, even with the other choice ($\theta-\lambda$ is an integer multiple of $\pi$) though the meaning of `time reflection' become obscure then. Moreover, the classical counterpart of the `parity' operation on the complex $X$-plane $\mathcal{P}: X \rightarrow -X^*$ naturally realizes the `left-right' reflection mentioned in Ref. \cite{Bender:2007nj}.}

{The eigenstates of $\hat{H}=(\hat{P}^2+\hat{X}^2)/2$ can be represented in $X$-space. Being pseudo-Hermitian, $\hat{X}^\sharp=\hat{X}$ has real spectrum and its eigenstates $\vert X\rangle$ are `orthonormal' with respect to $\eta$-inner product, that is, $\langle X\vert \eta\vert X'\rangle=\delta(X-X')$. This result leads to the completion relation; $\int\mathrm{d}X\vert X\rangle\langle X\vert\eta=\hat{I}$. In $X$-space representation, $b$-vacuum $\vert 0\rangle_b$, being annihilated by $\hat{b}=(\hat{X}+i\hat{P})/\sqrt{2}$, satisfies
\begin{equation}
\langle X\vert\eta\hat{b}\vert 0\rangle_b=\frac{1}{\sqrt{2}}\left( X+\partial_X\right)\langle X\vert\eta\vert 0\rangle_b=0.
\end{equation}
Therefore, the eigenfunctions $\langle X\vert\eta\vert n\rangle_b$ are given in terms of Hermite polynomial as
\begin{equation}
\langle X\vert\eta\vert n\rangle_b=\frac{e^{-\frac{X^2}{2}}}{\sqrt{2^n n!\sqrt{\pi}}}H_n(X).
\end{equation} }

{Regarding the probability density function, defined to be real {\it ab initio}, one can use two different well-defined representations. The relations
\begin{equation}
\frac{\int\!\mathrm{d}x \langle\psi\vert x\rangle\!\langle x\vert\psi\rangle}{\langle\psi\vert\psi\rangle}\!=\!\frac{\int\!\mathrm{d}X \langle\psi\vert \eta\vert X\rangle\!\langle X\vert\eta\vert\psi\rangle}{\langle\psi\vert\psi\rangle_{\eta}}\!=\!1
\end{equation}
suggest that the integrands (including the denominator) be the probability density functions in $x$-space and in $X$-space, respectively. 
}

{It is the expectation value of the position operator $\hat{x}$ that is the complex coordinate we meet in the non-Hermitian system. Exploiting Eqs. \eqref{repre} and \eqref{ladderb}, one can obtain the operator relations,
\begin{eqnarray}
\hat{x}&=&\hat{X} \cos{\theta}\,+\hat{P}\sin{\theta}\,+z^*\hat{I},  \nonumber\\
\hat{p}&=&-\hat{X} \sin{\theta}\,+\hat{P}\cos{\theta}\,+iz^*\hat{I}. 
\end{eqnarray}
The operators $\hat{x}$ and $\hat{p}$, though Hermitian with respect to $L^2$-inner product, have complex expectation values for any state $\vert \psi\rangle$ if we use $\eta$-inner product. For example in the expectation value $\langle\psi\vert\eta\hat{x}\vert\psi\rangle=\langle\psi\vert\eta\hat{X}\vert\psi\rangle \cos{\theta}\,+\langle\psi\vert\eta\hat{P}\vert\psi\rangle+z^*$,  the parts $\langle\psi\vert\eta\hat{X}\vert\psi\rangle$ and $\langle\psi\vert\eta\hat{P}\vert\psi\rangle$ are real with $\hat{X}$ and $\hat{P}$ pseudo-Hermitian, but $z^*$ is complex valued in general. The imaginary part $\Im(\langle\psi\vert\eta\hat{x}\vert\psi\rangle)=\Im(z^*)$ looks like an `order parameter' signaling the non-Hermiticity. Employing $L^2$-inner product, we can make the value $\langle\psi\vert\hat{x}\vert\psi\rangle$ real, but then the Hamiltonian will be invalidated, being complex valued.
}

{As regards the measurement, a proper choice of the inner product should be prior to determining the `physical' observables. Dirac-von Neumann axiom of measurement underpins quantum mechanics. In the conventional Hermitian system, we need not worry much about the inner-product despite that it is an essential component of Hilbert space. In the non-Hermitian system, we have to devise a new inner product that renders the Hamiltonian Hermitian. However, the operators $\hat{x}$ and $\hat{p}$ will be non-Hermitian in the new inner product, making them `unphysical' in the Dirac-von Neumann sense. }

{In conclusion, the entire information on the `complex position' $\hat{x}$ for a state is attainable, even though the operator is not pseudo-Hermitian. 
In this letter we constructed the `physical' position $X$ and `physical' momentum $P$ explicitly and found their relations with the `unphysical' counterparts, $\hat{x}$ and $\hat{p}$. From the relations we note that $\hat{x}-\Re{(z^*)}\hat{I}$ and $\hat{p}-i\Im{(z^*)}\hat{I}$ are pseudo-Hermitian, thus `physical'. One can always prepare for a state in an eigenstate of these pseudo-Hermitian operators. Therefore the pseudo-Hermitian part of the position operator $\hat{x}$ is definitely observable.  Its anti-pseudo Hermitian part is unobservable in the experiment, but it just reads the non-Hermiticity parameter $\Im(z^*)$. This part remains the same under the time flow. }

{It is interesting to see that the operator $\hat{x}-\langle\hat{x}\rangle_\eta\hat{I}$ is pseudo-Hermitian, thus, observable. The same is true for $\hat{p}$. This implies that albeit $\hat{x}$ and $\hat{p}$ are unobservable, their corresponding uncertainties are observable. Notably, they satisfy the uncertainty principle. }

\begin{acknowledgements}
The author thanks Seok Ho Song who first drew his attention to $\mathcal{PT}$-symmetric system, and Youngsun Choi, and Jae Woong Yoon for stimulating discussions on its application in optics area.
\end{acknowledgements}

\end{document}